\documentstyle[12pt,a4]{article}
\begin{document}
 
 \begin{center}
{\Large\bf A two-dimensional integrable axionic $\sigma$-model
\vskip 0.25 truecm
         and T-duality}
\end{center}
\vskip 2.5 truecm
\centerline{
{\large J\'anos Balog}}\vskip1ex
\centerline{Research 
Institute for Particle and Nuclear Physics,}
\centerline{Hungarian Academy of Sciences,}
\centerline{H-1525 Budapest 114, P.O.B. 49, Hungary} \vskip1ex
\centerline{ {\large P\'eter Forg\'acs}}
\vskip1ex
\centerline{Laboratoire de Math\'ematiques et Physique Th\'eorique}
\centerline{Universit\'e de Tours}
\centerline{Parc de Grandmont, 37200 Tours, France}\vskip1ex
\centerline{{\large L\'aszl\'o Palla}}\vskip1ex
\centerline{Institute for Theoretical Physics}
\centerline{E\"otv\"os University}
\centerline{H-1117 Budapest, P\'azm\'any P.\ s\'et\'any 1A, Hungary}

\vskip 3.20 true cm
\centerline{\bf Abstract}
An $S$-matrix is proposed for the two dimensional
O(3) $\sigma$-model with a dynamical $\theta$-term
(axion model).
Exploiting an Abelian T-duality transformation connecting the axion 
model to an integrable SU(2)$\times$U(1) symmetric principal
$\sigma$-model, strong evidence is presented 
for the correctness of the proposed $S$-matrix by
comparing the perturbatively
calculated free energies with the ones based on the Thermodynamical
Bethe Ansatz. This T-duality transformation also 
leads to a new Lax-pair for both models. 
The quantum non-integrability of
the O(3) $\sigma$-model with a {\sl constant} $\theta$-term,
in contradistinction to the axion model,
is illustrated by calculating
the $2\rightarrow3$ particle production amplitude
to lowest order in $\theta$.
\vfill
\eject

 
%
\newcommand{\eff}{\lambda_{\scriptscriptstyle {\rm eff}}}
\newcommand{\bee}{\begin{equation}}
\newcommand{\be}{\begin{equation}}
\newcommand{\ee}{\end{equation}}
\newcommand{\ba}{\begin{array}}
\newcommand{\ea}{\end{array}}
\newcommand{\bea}{\begin{eqnarray}}
\newcommand{\eea}{\end{eqnarray}}
\newcommand{\e}{\epsilon_+(i\kappa)}
\newcommand{\eps}{\epsilon}
\newcommand{\pa}{\partial}
\newcommand{\lb}{\lbrack}
\newcommand{\Se}{S_{\rm eff}}
\newcommand{\rb}{\rbrack}
\newcommand{\de}{\delta}
\newcommand{\th}{\theta}
\newcommand{\ka}{\kappa}
\newcommand{\al}{\alpha}
\newcommand{\si}{\sigma}
\newcommand{\vp}{\varphi}
\newcommand{\g}{\gamma}
\newcommand{\om}{\omega}
\newcommand{\pr}{\prime}
\newcommand{\gbb}{\bar{g}}
\newcommand{\gb}{\overline g}
\newcommand{\nb}{\overline N}
\newcommand{\MSb}{{\overline {\rm MS}}}
\newcommand{\lnh}{\ln(h^2/\Lambda^2)}
\newcommand{\df}{\delta f(h)}
\newcommand{\h}{{1\over2}}
\newcommand{\R}{m/\Lambda}
\newcommand{\tl}{\tilde{\lambda}}
\newcommand{\tg}{\tilde{g}}
\newcommand{\abschnitt}[1]{\par \noindent {\large {\bf {#1}}} \par}
\newcommand{\subabschnitt}[1]{\par \noindent
                                          {\normalsize {\it {#1}}}
\par}
\newcommand{\eqalign}[1]{
\null \,\vcenter {\openup \jot \ialign {\strut \hfil $\displaystyle {
##}$&$\displaystyle {{}##}$\hfil \crcr #1\crcr }}\,}
\newcommand{\skipp}[1]{\mbox{\hspace{#1 ex}}}
 
%
%
%
%
\newcommand\dsl{\,\raise.15ex\hbox{/}\mkern-13.5mu D}
\newcommand\delsl{\raise.15ex\hbox{/}\kern-.57em\partial}
\newcommand\Ksl{\hbox{/\kern-.6000em\rm K}}
\newcommand\Asl{\hbox{/\kern-.6500em \rm A}}
\newcommand\Dsl{\hbox{/\kern-.6000em\rm D}} 
\newcommand\Qsl{\hbox{/\kern-.6000em\rm Q}}
\newcommand\gradsl{\hbox{/\kern-.6500em$\nabla$}}

\pagestyle{plain}
\setcounter{page}{1}


In this paper we shall study the following two dimensional
$\sigma$-model described by the 
Lagrangian\footnote{We use the following conventions. For
a vector $v$ in two-dimensional Minkowski space 
$v^\mu v_\mu=v_+v_-$ where $v_\pm=v_0\pm v_1$. 
The antisymmetric tensor is defined by $\epsilon^{01}=1$,
$\tau^a=\sigma^a/2$
with $\sigma^a$ being the standard Pauli-matrices satisfying 
$\sigma^a\sigma^b=\delta^{ab}+i\epsilon^{abc}\sigma^c$.}:
\bee
{\cal L}=
{1\over2\tilde\lambda}\partial_\mu n^a \partial^\mu n^a+
{\tilde\lambda\over32\pi^2(1+\tilde g)}
\partial_\mu\theta\partial^\mu\theta+
{\theta\over8\pi}\epsilon^{\mu\nu}\epsilon^{abc}
n^a\partial_\mu n^b\partial_\nu n^c\,,
\label{Lagax}
\ee
where $n^an^a=1$. In fact (\ref{Lagax})
is an O(3) non-linear $\sigma$-model, 
coupled to a scalar field, $\theta$, 
(whose normalization has been chosen for later convenience) 
 through the Hopf term. This latter is proportional 
to the topological current of the O(3) model and with the  
normalization chosen in (\ref{Lagax}) its space-time integral 
(after a Wick rotation) yields the topological charge, 
which can take integer values only. This implies that the 
variable $\theta$ is actually an angle,
taking its values between $0$ and $2\pi$. This observation will play an 
important role in all our considerations.

We shall refer to (\ref{Lagax}) as the `axion model' since it
can be thought of as the
O(3) nonlinear $\sigma$-model with a dynamical $\theta$-term
which can be regarded 
as a two-dimensional analogue of its 
phenomenologically important four-dimensional counterpart 
\cite{axions}.
The following (heuristic) consideration might be useful to gain some
insight into the physics of the axion model.
Let us integrate out the O(3) fields, 
$n^a$, in some generating functional of the theory (\ref{Lagax}). 
This way one would obtain a non-vanishing effective potential for 
the $\theta$ field.
Because $\theta$ is 
$2\pi$-periodic the effective potential must be also 
periodic. The effective theory of the $\theta$ field
is therefore expected to be similar to the 
Sine-Gordon model, with a periodic potential and corresponding 
topological current $K_\mu=\epsilon_{\mu\nu}
\partial^\nu\theta/2\pi$. 
Note that $\theta$
being an angular variable makes it difficult to integrate it out in
the functional integral in spite of ${\cal L}$ being only quadractic
in $\theta$.

The axion model 
belongs to a family of (classically) integrable
two-dimensional non-linear $\sigma$-models with an
O(3) symmetry discovered in Ref.\ \cite{class}.
All of the models of Ref.\ \cite{class} have target spaces with
non-vanishing torsion (in addition to the metric tensor field) 
but the axion model is especially simple, as its
torsion is constant. 

It has already been observed in Ref.\ \cite{bfhp} that the axion model 
(\ref{Lagax}) can be mapped to a one parametric deformation of the 
SU(2) principal $\sigma$-model by
an Abelian T-duality transformation. 
This latter (torsionless) model has an SU$_{\rm L}(2)\otimes$U$_{\rm R}$(1)
symmetry and recently there has been some revival of interest in it
 \cite{Fateev,BF}.
Its Lagrangian can be written as: 
\bee
{\cal L}_\Sigma=-{1\over2\lambda}\Big\{L^a_\mu\,L^{a\mu}+
gL^3_\mu\,L^{3\mu}\Big\}\,,
\label{Lagdef}
\ee
where
\bee
L_\mu=G^{-1}\partial_\mu G=\tau^aL^a_\mu\,,
\label{L}
\ee
and $g$ is the parameter of deformation.
The theory (\ref{Lagdef}) is known to be integrable classically
\cite{Cherednik}
and there is little doubt that it is also quantum-integrable 
\cite{Fateev,BF,Kirillov}.
Its spectrum contains two massive doublets (kinks)
whose scattering is described by the tensor product of
an SU(2)$\times$U(1) symmetric solution of the bootstrap $S$-matrix
equations:
\bee
S(\theta)=S^{(\infty)}(\theta)\otimes S^{(p)}(\theta)\,,
\label{Smatrix}
\ee 
where $S^{(p)}(\theta)$ denotes the Sine-Gordon (SG) $S$-matrix.
Depending on the value of the parameter, $p$,
in addition to the kinks there are also
some bound states (breathers) 
in the spectrum transforming as $3+1$ under SU(2).

In the following we shall show that 
the somewhat unexpected identification between the axion model and 
the deformed principal $\sigma$-model through a T-duality transformation
allows us to learn more about both of them.
Assuming the validity of 
the duality transformation at the {\sl quantum level} 
between the two theories 
implies the absence of particle production in the axion 
model (\ref{Lagax}), and also that its factorized scattering theory
is given by the $S$-matrix of Eq.\ (\ref{Smatrix}).

The proposed  quantum integrability
of the axion model might seem  somewhat surprising, as
it is generally believed that the O(3) model with a {\sl constant} 
$\theta$-term is {\sl not quantum integrable},
except for the special value $\theta=\pi$ \cite{SMpi}
(despite the fact that the $\theta$-term, being a total derivative, 
does not change the classical physics of the model).
We now show that in the framework of the form-factor bootstrap 
approach the $\theta$ term mediates 
particle production in the O(3) $\sigma$-model, indeed. 
To lowest order in $\theta$ the $2\to3$ 
particle production amplitude can be written as
\bea
&&\langle p,b;p^\prime,b^\prime;p^{\prime\prime},b^{\prime\prime}\vert
q,a;q^\prime,a^\prime\rangle_{(\theta)}=(2\pi)^2\,i\,\theta\,\delta^{(2)}(
p+p^\prime+p^{\prime\prime}-q-q^\prime)\nonumber\\
&&\qquad\qquad \cdot\,
\langle p,b;p^\prime,b^\prime;p^{\prime\prime},b^{\prime\prime}\vert
T(0)\vert q,a;q^\prime,a^\prime\rangle_{(0)}+
{\cal O}(\theta^2)\,,
\label{amplitude}
\eea
where in the first line the amplitude is in the 
O(3) model with a $\theta$-term, while in the second line the
matrix element of the topological charge density operator $T$ is to be
calculated in the original O(3) $\sigma$-model (with $\theta=0$).
In other words we simply apply perturbation theory in $\theta$.
Let us now consider the following simplified
kinematical configuration: the incoming particles have momenta $q_1=Q$
and $q_1^\prime=-Q$, whereas the produced (outgoing) three particles
have momenta $p_1=Q^\prime$, $p_1^\prime=0$ and $p_1^{\prime\prime}=
-Q^\prime$ respectively. Here $Q^\prime$ can easily be expressed in
terms of $Q$ and the kink mass $M$ using energy conservation. For 
large $Q$, using the results of Ref.\ \cite{BN}, we find
\be
\langle p,b;p^\prime,b^\prime;p^{\prime\prime},b^{\prime\prime}\vert
T(0)\vert q,a;q^\prime,a^\prime\rangle_{(0)}\approx
{\pi}^{5\over2}\frac{Q^2}{\ln^3Q/M}\left(\epsilon^{a^\prime ba}\delta^{
b^\prime b^{\prime\prime}}-
\epsilon^{b^{\prime\prime} ba}\delta^{
b^\prime a^\prime}\right)\,.
\label{asy}
\ee
Eq.\ (\ref{asy}) shows that already to first order in $\theta$, the
$2\to3$ particle production amplitude is different from zero.
Thus at least for small values of $\theta$, the introduction of this term 
destroys quantum integrability of the O(3) $\sigma$-model, indeed.

To exhibit now the classical T-duality transformation
between the two models \cite{bfhp}, we introduce the parametrization
\bee
n^1=\sin\vartheta\sin\varphi\,,\quad
n^2=\sin\vartheta\cos\varphi\,,\quad
n^3=\cos\vartheta\,,\qquad
\theta=-{4\pi\over\tilde\lambda}\sqrt{1+\tilde g}\,\chi\,,
\label{para_ax}
\ee
in terms of which the Lagrangian (\ref{Lagax}) (after
an integration by parts) becomes
\bee
{\cal L}={1\over2\tilde\lambda}\Big\{
\partial_\mu\vartheta\partial^\mu\vartheta
+\sin^2\vartheta\partial_\mu\varphi\partial^\mu\varphi
+\partial_\mu\chi\partial^\mu\chi
+2\sqrt{1+\tilde g}\cos\vartheta\epsilon^{\mu\nu}
\partial_\mu\chi\partial_\nu\varphi\Big\}\,.
\label{Lagax_par}
\ee
We now perform an Abelian T-duality transformation \cite{Busch}
with respect to the variable $\chi$, which corresponds to the
canonical transformation \cite{Alvarez}
\bee
\chi^\prime=-{\tilde\lambda\over\sqrt{1+\tilde g}}p_\alpha\qquad\qquad
p_\chi=-{\sqrt{1+\tilde g}\over\tilde\lambda}\alpha^\prime\,,
\label{cano}
\ee
where (and in the following) $p_\chi$ resp.\  $p_\alpha$ denote the 
canonical momenta conjugate to $\chi$ resp.\ to its `dual' 
$\alpha$. In terms of these new variables
the dual Lagrangian turns out to be:
\bee
{\cal L}_\Sigma={1\over2\tilde\lambda}\Big\{
\partial_\mu\vartheta\partial^\mu\vartheta
+(1+\tilde g\cos^2\vartheta)\partial_\mu\varphi\partial^\mu\varphi
+(1+\tilde g)\left[\partial_\mu\alpha\partial^\mu\alpha
+2\cos\vartheta
\partial_\mu\alpha\partial^\mu\varphi\right]\Big\}\,,
\label{Lagdef_par}
\ee 
which is nothing but 
the Lagrangian (\ref{Lagdef}), when parametrizing the SU(2) valued field, $G$,
by the Euler angles
\bee
G=e^{i\varphi\tau^3}\,e^{i\vartheta\tau^1}\,
e^{i\alpha\tau^3}\,,
\label{G}
\ee
and taking into account the relations at the classical level
between the couplings:
\bee\label{couplrel}
\tilde\lambda =\lambda\,,\quad 
\quad \tilde g=g\,.
\ee
The observation, that the axionic model is the T dual of 
${\cal L}_\Sigma$ also explains why $\theta $ is an angular variable. 
Indeed
it has been shown in Ref.\ \cite{alvtop} that in case of the principal 
$\sigma$-model
($g=0$) the Abelian T duality (\ref{cano}) maps its target space 
($S^3$) into 
$S^2\times S^1$. The arguments of
Ref.\ \cite{alvtop} can be easily applied to the present case with
$g>-1$, and it is clear that in Eq.\ (\ref{Lagax}) $n^a$ parametrize the
$S^2$ and $\theta $ parametrizes the $S^1$.  

In fact the equations of motion of
both models (\ref{Lagax}) and (\ref{Lagdef}) are known to
admit a Lax representation indicating their (classical)
integrability \cite{class,Cherednik}.
Indeed, introducing the matrix valued
current
\bee
I_\mu={\tilde\lambda\over8\pi}n\epsilon_{\mu\nu}\partial^\nu\theta-
{\sqrt{\tilde g}\over2}\epsilon_{\mu\nu}\partial^\nu n+{1\over2}
n\partial_\mu n\,,
\label{I}
\ee
where $n=in^a\sigma^a$, the equations of motion of (\ref{Lagax}) 
can be written as:
\bee
\partial^\mu I_\mu=0\,,\qquad\qquad
\partial_\mu I_\nu-\partial_\nu I_\mu=[I_\mu,I_\nu]\,.
\label{LaxI}
\ee
The standard form (\ref{LaxI}) of the equations of motion
allows for the introduction of a Lax pair
\bee
U_\pm={1\over1\pm\omega}\,I_\pm\,,
\label{LaxU}\ee
satisfying the zero curvature equation
\bee
\partial_\mu U_\nu-\partial_\nu U_\mu=[U_\mu,U_\nu]\,,
\label{zeroU}
\ee
for all values of the spectral parameter $\omega$. 
The current, $I_\mu$, is closely related to the matrix valued
Noether current, ${\cal N}_\mu=-i\tau^a{\cal N}^a_\mu$,
defined by $\delta{\cal L}=\partial^\mu
\varepsilon^a{\cal N}^a_\mu$ corresponding to the symmetry
transformation $\delta n^a=\epsilon^{abc}\varepsilon^bn^c$:
\bee
I_\mu=\tilde\lambda{\cal N}_\mu+\epsilon_{\mu\nu}\partial^\nu T\,,
\qquad\qquad T=\Big({\tilde\lambda\over8\pi}\theta-{\sqrt{\tilde g}
\over2}\Big)n\,.
\label{IN}
\ee
The fact that, apart from a trivially conserved piece,
$I_\mu$ can be identified with the Noether
current of the manifest O(3) symmetry of the Lagrangian
explains only the first equation in Eq.\ (\ref{LaxI}).
The trivially conserved part of $I_\mu$ is essential that
the zero curvature equation be also satisfied.

The equations of motion of the deformed principal model (\ref{Lagdef})
can be written entirely in terms of the current $L_\mu$ as
\bee
\partial^\mu L^3_\mu=0\,,\quad
\partial^\mu L^1_\mu=-igL^{2\mu}L^3_\mu\,,\quad
\partial^\mu L^2_\mu=igL^{1\mu}L^3_\mu\,.
\label{EOM}
\ee
It is known that this system can be put to the Lax form 
\cite{Cherednik,Kirillov}, i.e.\ there is a spectral parameter dependent 
current, $V_\mu=\tau^aV^a_\mu$, satisfying the zero 
curvature equation (\ref{zeroU}). This current can be written as: 
\bee
V^{1,2}_\pm=\alpha_\pm\,L^{1,2}_\pm\,,\qquad\qquad
V^3_\pm=a_\pm\,L^3_\pm\,,
\label{V}
\ee
where
\bee
\alpha_\pm=-{4+g\omega^2\over4-g\omega^2\pm4\omega}\,,\qquad
a_\pm=-{4-g\omega^2\mp4g\omega\over4-g\omega^2\pm4\omega}\,.
\label{alphapm}
\ee

We can now use the classical T-duality transformation (\ref{cano}) to map the
linear system of the axion model (\ref{LaxI}) to
a new Lax pair for the deformed $\sigma$-model (\ref{Lagdef}). 
It is given by Eq.\ (\ref{LaxU}), where the current, $I_\mu$, has to be
replaced by 
\bee
\hat I_\mu=\partial_\mu G\,G^{-1}+{g}
\Big(G\tau^3G^{-1}\Big)
L^3_\mu-{i\sqrt{g}}\epsilon_{\mu\nu}\partial^\nu
\Big(G\tau^3G^{-1}\Big)\,.
\label{dualI}
\ee
Eq.\ (\ref{dualI}) is obtained from (\ref{I}) by the T-duality
transformation (\ref{cano}). $\hat I_\mu$ is related to the Noether
current $\hat{\cal N}_\mu$, corresponding to the manifest
symmetry $\delta G=-i\varepsilon^a\tau^a G$ of
(\ref{Lagdef}) and can be written analogously to ${\cal N}_\mu$:
\bee
\hat I_\mu=\lambda\hat{\cal N}_\mu+
\epsilon_{\mu\nu}\partial^\nu \hat T\,,
\qquad\qquad \hat T=-{i\sqrt{g}}
\, G\tau^3G^{-1}\,.
\label{IhatN}
\ee
It is clear that the new Lax pair (\ref{LaxU}) and 
the \lq old' one, (\ref{V}), cannot be related by
a gauge transformation since they have different pole
structures as functions of the spectral variable, $\omega$.
In the $g\to-1$ limit, 
the axion model reduces to the original 
O(3) $\sigma$-model (decoupled from the $\theta$ field), 
and the Lax pair (\ref{LaxU})
becomes equivalent to that of Ref.\ \cite{Byt}, where it has been pointed out
that the corresponding $\hat I_\mu$'s are {\sl ultralocal} currents.
We note
that the Lax pairs (\ref{LaxU}) and (\ref{V}) correspond to (different)
 deformations
of the usual Lax pairs of the principal chiral $\sigma$-model, 
linear in $\partial_\mu GG^{-1}$ respectively $G^{-1}\partial_\mu G$.

Next we carry out a standard test on the proposed $S$-matrix (\ref{Smatrix}) 
for the axion model
by comparing its (zero temperature) free energy obtained from the
Thermodynamical Bethe Ansatz (TBA) and in weak coupling perturbation theory
(PT)
\cite{history}. For the deformed $\sigma$-model (\ref{Lagdef}) this comparison
has been done in Ref.\ \cite{BF} where complete consistency has been found
between the results of PT and of the TBA.
For the axion model it is sufficient to compute the free energy in PT 
as the results of the TBA can be literally taken over from Ref.\ \cite{BF}.
In the present case 
one obtains as a bonus, a further nontrivial check on the quantum equivalence
between the axion and the deformed $\sigma$-model, hence also on the
validity of the T-duality transformation at the quantum level. 
Up to now when quantum equivalence between dually related 
models has been tested, mostly $\beta$-functions have been 
compared. The fact that the higher coefficients of the $\beta$-functions
are scheme dependent makes such a comparison more
difficult and less conclusive. 

The equivalence of the $\beta$-functions is certainly a necessary condition 
for the validity of quantum T-duality.
At one loop order the $\beta$-functions of
the couplings, $\beta_\lambda$, $\beta_g$ 
and $\beta_{\tilde{\lambda}}$, $\beta_{\tilde{g}}$ are simply 
obtained from each 
other by the classical relation (\ref{couplrel}).
At two loops, however, it has been found in \cite{bfhp} 
that using the background field method and dimensional regularization
the following perturbative redefinition of the couplings  
\bee\label{psicoupl}
\tilde\lambda=\lambda+{\lambda^2\over4\pi}(1+g)\,,\quad
\tilde g=g+{\lambda\over4\pi}(1+g)^2\,,
\ee
(i.e.\ a change of scheme)
is induced by the T-duality transformation. Taking into account
Eqs.\ (\ref{psicoupl}) the two loop
$\beta$-functions of the two models turn out to be equivalent.
Alternatively, introducing a renormalization group (RG)
invariant combination of the two couplings:
\bee
p=2\pi\lim_{t\rightarrow\infty}{1+g(t)\over\lambda(t)}\,,
\qquad
\tilde p=2\pi\lim_{t\rightarrow\infty} 
{1+\tilde g(t)\over\tilde\lambda(t)}\,,
\label{g}
\ee
where $t\propto\ln h$, one finds $p=\tilde p$ up to two loops \cite{bfhp}.
It is important to note that the RG invariant quantity (\ref{g})
can be consistently {\sl identified} with the parameter $p$ in the 
$S$ matrix (\ref{Smatrix}).
Let us introduce an effective $\beta$-function for $\lambda(t)$ by 
$\beta_{\scriptscriptstyle{\rm eff}}(\lambda,p)
=\beta_\lambda(\lambda,\Gamma(\lambda,p))$,
expressing 
$g(t)$, in terms of the running coupling, $\lambda(t)$,
and $p$ as
$g(t)=\Gamma(\lambda(t)\,,p)$.
Using the perturbative result for $\Gamma(\lambda(t),p)$ \cite{BF}
 one finds
\bee
\beta_{\scriptscriptstyle{\rm eff}}(\lambda,p)=
\beta_{\scriptscriptstyle{\rm eff}}(\tilde\lambda,\tilde p)
=-{\lambda^2\over2\pi}+
{p-2\over8\pi^2}\lambda^3+\cdots\,.
\label{betaeff1}
\ee
Thus as far as coupling constant renormalization is concerned,
the two models are equivalent, both 
are asymptotically free, and the actual value of $p$ effects only the 
two loop coefficient.

The classical free energy density is obtained
by minimizing the Legendre transform of the Hamiltonian density 
coupled to some conserved currents
\be\label{hamgen}
\hat{\cal H}={\cal H}_0-h_iJ^i_0\,,\qquad 
\hat{H}=\int dx\hat{\cal H}=H-h_iQ_i\,.
\ee 
Since the axion field, $\theta$, is actually an angle, 
its winding number (topological charge) can be non trivial.
Therefore we present here the Legendre transformation
of the modified Hamiltonian (\ref{hamgen}) for a rather general 
case.

Let us consider a general sigma model with torsion
\be\label{Laggen}
{\cal L}_0=\frac{1}{2}g_{AB}\pa^\mu X^A\pa_\mu X^B+
\frac{1}{2}b_{AB}\epsilon^{\mu\nu}\pa_\mu X^A\pa_\nu X^B,
\ee
and the following Ansatz for a set of conserved currents
\be\label{curr} 
J_\mu^i={\cal C}_A^i(X)\pa_\mu X^A+\epsilon_\mu^{\ \nu}{\cal
B}_A^i(X)\pa_\nu X^A\,,
\ee
sufficiently general to include topological currents.
The Legendre transformation of (\ref{hamgen}) yields the
Lagrangian of the modified model which can be written as
\be
\hat{\cal L}={\cal L}_0+h^iJ^i_0+\frac{1}{2}h^ih^j{\cal C}_A^i{\cal
C}^{A j}\,.
\ee
In fact $\hat{\cal L}$ can be obtained by gauging ${\cal L}_0$ i.e.\ 
by the substitution 
\be\label{gauging}
\pa_\mu X^A\to\pa_\mu X^A +h^i\delta_{\mu 0}{\cal C}^{iA}
\ee
when the antisymmetric field $b_{AB}$ is invariant (without compensating
gauge transformation)
under the symmetry transformation generated by the conserved currents
(\ref{curr}).

Below we also give a class of classical ground states (around which $\hat{\cal L}$
is to be expanded) assuming that the metric, the
antisymmetric tensor field and the quantities characterizing
the currents are independent of a set of coordinates, $\theta^\alpha$,
corresponding to the splitting $X^A=(y^k,\theta^\alpha)$:  
\be
g_{AB}=g_{AB}(y),\quad b_{AB}=b_{AB}(y),\quad {\cal C}_A^i={\cal
C}_A^i(y),\quad {\cal B}_A^i={\cal B}_A^i(y)\,.
\ee
In this case the ground state is characterized by constant $y^k$-s and
constant ${\theta^\prime}^\alpha$-s $y^k\equiv y_0^k$, 
${\theta^\prime}^\alpha\equiv {\theta^\prime}_0^\alpha$, where
the $y_0^k$-s stand for the extrema of 
\be\label{heff}
H_{\rm eff}=-\frac{1}{2}h^ih^j(g_{AB}{\cal C}^{Ai} {\cal C}^{Bj}+
{\cal B}_\alpha^i {\cal B}_\beta^j (\gamma^{-1})^{\alpha\beta})\,,
\ee
and 
\be\label{tetder}
{\theta^\prime}_0^\alpha=(\gamma^{-1})^{\alpha\beta}{\cal
B}_\beta^i(y_0)h^i.
\ee
In (\ref{heff}-\ref{tetder}) 
$\gamma_{\alpha\beta}$ denotes the restriction of $g_{AB}$ to the 
submanifold coordinatized by 
$\theta^\alpha$.

The axion model has a `manifest' (i.e.\ up to a total derivative) 
SU(2)$\times{\rm U}_\theta(1)$ symmetry,  
where the ${\rm U}_\theta(1)$ subgroup is generated by the shift 
$\theta\rightarrow\theta+{\rm const}$.
It is very important to note that 
although the SU(2) symmetry of the Lagrangian (\ref{Lagax}) corresponds to 
that of the $S$-matrix, the `manifest' ${\rm U}_\theta(1)$ symmetry 
{\sl cannot be identified} with the corresponding one of the $S$-matrix 
(\ref{Smatrix}).
Here the duality transformation
provides the clue; the corresponding $\tilde {\rm U}_{\rm R}$(1) symmetry of the axion 
model is actually 
the image of the manifest U$_{\rm R}$(1) symmetry of (\ref{Lagdef}) 
under the duality transformation, thus it is generated by the {\sl
topological current} of the axion field. 

Corresponding to the SU$_{\rm L}(2)\otimes$U$_{\rm R}$(1) symmetry of the 
deformed $\sigma$-model there are two Noether 
charges, $Q_{\rm L}$ resp.\ $Q_{\rm R}$, associated to the 
U$_{\rm L}$(1) resp.\ U$_{\rm R}$(1) subgroups.
Introducing two chemical potentials coupled to the $Q_{\rm L}$ resp.\ $Q_{\rm R}$,
charges the Hamiltonian (\ref{hamgen}) takes the form: 
$ H=H_{\Sigma} - h_{\rm L}Q_{\rm L}-h_{\rm R}Q_{\rm R}\,.$
Then 
one can distinguish between three different types of finite density ground states:
LEFT with $h_{\rm L}>0,$ and $h_{\rm R}=0,$ 
RIGHT with $h_{\rm L}=0,$ and $h_{\rm R}>0,$
and DIAG where $h_{\rm L},h_{\rm R}>0$. As found in \cite{BF} the RIGHT case
is obtained from DIAG by letting $h_{\rm L}=0$ in the {\sl final} results.

We compute below the corresponding ground 
state energies to one loop order in the axion model (\ref{Lagax}),
starting  with the LEFT case first.
With the Euler angle parametrization of $G$ (\ref{G}) 
the U$_{\rm L}(1)$ transformation,
$G\mapsto e^{i\kappa\tau^3}G$ of the deformed $\sigma$-model (\ref{Lagdef}) 
acts as a simple shift, $\varphi (x)\mapsto\varphi(x)+\kappa$.
The corresponding Noether charge, $Q_{\rm L}$, and its image under
the T-duality transformation, $\tilde Q_{\rm L}$, are simply
 $$Q_{\rm L}=\int dxp_\varphi\,,\quad
 \tilde{Q}_{\rm L}=\int dx
\tilde{p}_\varphi\,,\quad
 {\rm where}\quad 
 p_\varphi=\frac{\pa {\cal L}_\Sigma}{\pa\dot{\varphi}}\,,\;
\tilde{p}_\varphi=\frac{\pa {\cal L}}{\pa\dot{\varphi}}\,,
$$
since the canonical transformation
implementing the T-duality mapping (\ref{cano}) effects only 
$p_\alpha$, $\chi^\prime$, 
$\alpha^\prime $ and $p_\chi$, 
leaving the other fields,
$\varphi $, $\vartheta $, $p_\varphi$, $p_\vartheta$, unchanged. 

Since in the LEFT case the $b_{AB}$ field in Eq.\ (\ref{Lagax}) is invariant,
one can simply `gauge' the Lagrangian of the
axion model in an external ($h_{\rm L}$) field (see Eq.\ (\ref{gauging}).  
The classical ground state is found to be
 $\varphi\equiv\chi\equiv 0$, $\vartheta\equiv
{\pi}/2$.
(The corresponding solution of the deformed $\sigma$-model 
is given by $\varphi\equiv\alpha\equiv 0$, $\vartheta\equiv
{\pi}/2$.) 

Expanding the (Euclidean) Lagrangian (after suitable rescalings, etc.)
we obtain
\be\label{kozlag}
\overline{\cal L}=-\frac{2h^2_{\rm L}}{\tilde{\lambda}_0}+\frac{1}{2}m{\cal M}m^T
+{\tt o}(\tilde{\lambda}), 
\ee
where 
\be\label{du0}
{\cal M}=\pmatrix{-\pa^2+4h_{\rm L}^2&0&2h_{\rm L}\sqrt{1+\tilde{g}_0}\,
\epsilon_{\mu 2}\pa_\mu\cr 0&-\pa^2&0\cr 
-2h_{\rm L}\sqrt{1+\tilde{g}_0}\,\epsilon_{\mu 2}\pa_\mu&0&-\pa^2\cr},
\ee  
and $m =(\vartheta, \varphi ,\chi )$. ($\tilde{\lambda}_0$,
$\tilde{g}_0$ denote the bare coupling and parameter of the axion/dual
model).    
 In Eq.~(\ref{du0}) we kept the $\epsilon $
tensor explicitly, as it requires a careful definition in
$n=2-\epsilon $ dimensions which we use to regularize the momentum
integrals. We adopt the definiton of \cite{Osb}, where this 
antisymmetric tensor corresponds to an almost complex structure:
$\epsilon_{\mu\nu}=-\epsilon_{\nu\mu}$,
$\epsilon_{\mu\nu}\epsilon_{\mu\sigma}=\delta_{\nu\sigma}$.
The one loop quantum corrections to the classical ground state
(the first term in Eq.\ (\ref{kozlag})) require the calculation
of a functional determinant, leading to
\be\label{det}
{\cal F}(h)=\frac{4h_{\rm L}^2}{n}\int\frac{d^np}{(2\pi )^n}
\frac{\tilde{p}_1^2-\tilde{g}_0\tilde{p}_2^2}{{\tilde{p}}^4+
4h_{\rm L}^2(\tilde{p}_1^2-
\tilde{g}_0\tilde{p}_2^2)}\,,\quad \tilde{p}_\mu=\epsilon_{\mu\nu}p_\nu\,.
\ee
To evaluate (\ref{det}) we apply the modified 
dimensional regularization of Ref.\ \cite{BF} 
as $\tilde{p}_2=\epsilon_{2\nu}p_\nu$ plays here a distinguished role
and it is kept as a one dimensional variable.
In fact for our purposes it is sufficient to calculate 
the {\sl difference} ${\cal F}(h)-{\cal F}_{\Sigma}(h)$,
where ${\cal F}_{\Sigma}(h)$ is the corresponding determinant
in the deformed $\sigma$-model (Eq.\ (3.12) in Ref.\ \cite{BF}).
Since both ${\cal F}(h)$ and ${\cal F}_{\Sigma}(h)$ are already the
first quantum corrections to the classical expressions we may set
$\tilde{g}=g$ (and make no distinction between bare and renormalized
$g$'s) when computing their difference to lowest order and
we end up with 
\be\label{F0dif}
{\cal F}(h)-{\cal F}_{\Sigma}(h)=
\frac{(2h_{\rm L})^n}{n}(1+g)\int\frac{d^nq}{(2\pi )^n}
\frac{(q_1^2-q_2^2)q^4}{N_1N_2}=\frac{(2h_{\rm L})^n}{n}(1+g)w(g),
\ee
where $N_1=q^4+q_1^2-gq_2^2$, $N_2=q^4+q_2^2-gq_1^2$. 
Although the integrand yielding $w(g)$ is 
antisymmetric under $q_1\leftrightarrow q_2$, the integral is
divergent by power counting for $n=2$, i.e.\ it must be
computed in $n=2-\epsilon$ dimensions.
Its derivative, $w^\prime (g)$, is, however, {\sl convergent} by power 
counting
and it has also an antisymmetric integrand, therefore
this latter may be evaluated in $n=2$ dimensions giving 
$w^\prime (g)\equiv 0$. Then to compute $w(g)$
one may choose e.g.\ the point $g=-1$:
\be\label{wme}
w(-1)=\int\frac{d^{n}q}{(2\pi )^n}
\frac{q_1^2-q_2^2}{(q^2+1)^2}=\frac{n-1-1}{n} \int\frac{d^nq}{(2\pi )^n}
\frac{q^2}{(q^2+1)^2}=-\frac{1}{4\pi},
\ee
where writing the second equality, we used that $q_1$ is $n-1$ dimensional,
while $q_2$ is a $1$ dimensional variable. 
From (\ref{wme}) one finds that after taking into account the change of the
renormalization scheme (\ref{psicoupl}), in PT the free
energy densities of the two models (\ref{Lagax}) and (\ref{Lagdef})
do indeed coincide for the LEFT case.
Recently this calculation  has been performed also at the two-loop level 
\cite{Karp}.

To discuss the RIGHT and DIAG cases we find it more convenient to 
use the parametrization of \cite{BF} for the SU(2) valued field, $G$: 
\bee
G={i\sigma^2\over\sqrt{1+\vert \Psi\vert^2}}\pmatrix{1&-\Psi^*\cr \Psi&1\cr}
\pmatrix{e^{-i\Phi}&0\cr0&e^{i\Phi}\cr}\,,
\label{para}
\ee
where $\Psi$ resp.\ $\Phi$ is a complex resp.\ a real scalar field.
Now $U_{\rm R}(1)$ acts as a
shift, $\Phi\mapsto\Phi +\kappa$, and then the corresponding
Noether charge of the deformed $\sigma$-model
is simply $Q_{\rm R}=\int dx p_\Phi$. 
$Q_{\rm L}$ is slightly more complicated
when expressed in terms of the canonical momenta
(as $e^{i\kappa\sigma^3}i\sigma^2=i\sigma^2e^{-i\kappa\sigma^3}$):
$
Q_{\rm L}=\int dx[p_\Phi+2i(p_\Psi\Psi-p_{\Psi^*}\Psi^*)]\,.
$

Using Buscher's rule \cite{Busch}, the Lagrangian
of the axion (dual) model now takes the form: 
\be
{\cal L}^d=\frac{\tl}{8(1+\tg)}(\pa_\mu
f)^2+\frac{2}{\tl}\frac{\pa_\mu\Psi\pa^\mu\Psi^*}{N^2}+\frac{1+\tg}{\tl}
\hat{\cal A}_\mu\hat{\cal A}^\mu
-\frac{i}{2}\epsilon^{01}(\dot{f}\hat{\cal A}_1-f^\prime \hat{\cal
A}_0)\,,
\ee
where $f$ is the dual to $\Phi$, and  
\be
\hat{\cal A}_\mu={\cal A}_{\Psi^*}\pa_\mu\Psi^*-{\cal
A}_\Psi\pa_\mu\Psi=\frac{1}{N}(\Psi\pa_\mu\Psi^*-\Psi^*\pa_\mu\Psi)\,,
\quad N=1+\vert\Psi\vert^2\,.
\ee
The canonical transformation connecting ${\cal L}_\Sigma$ and
${\cal L}^d$, maps $Q_{\rm R}$ and $Q_{\rm L}$  to
\be
\tilde{Q}_{\rm R}=-\int dx f^\prime\,,\quad
\tilde{Q}_{\rm L}=\int dx \left[-f^\prime +
2i(\tilde{p}_\Psi\Psi-\tilde{p}_{\Psi^*}\Psi^*)\right]\,,
\ee
i.e.\ $\tilde{Q}_{\rm R}$ and $\tilde{Q}_{\rm L}$ do indeed contain the
topological charge of the axion field (proportional to $f$).

Applying now the general framework, Eqs.\ (\ref{heff}-\ref{tetder}) 
to the present cases;
$i=$(L,R), $X^A=($$f$,$\Psi$,$\Psi^*$),
 with $\theta^\alpha=(f)$, $y^k=(\Psi ,\Psi^*)$. 
Using the explicit form of $\tilde{p}_\Psi$
and $\tilde{p}_{\Psi^*}$ one finds for $\tilde{J}_\mu^{\rm R,\ L}$:
\be
{\cal C}_A^{\rm R}\equiv 0,\qquad {\cal B}_A^{\rm R}=
 \left\{\begin{array}{ll}
 -1, \quad A=f\\
\ \ 0,\quad A=\Psi ,\ \Psi^*,
\end{array}\right.
\ee
\be
{\cal C}_A^{\rm L}=
 \left\{\begin{array}{lll}
 0, \quad A=f\\
 -{\cal N}{\cal A}_\Psi/{\tl},\quad A=\Psi\\
{\cal N}{\cal A}_{\Psi^*}/{\tl},\quad A=\Psi^*
\end{array}\right.
\qquad
{\cal B}_A^{\rm L}=
 \left\{\begin{array}{ll}
 -(1-2{\vert\Psi\vert^2}/{N}), \quad A=f\\
\ \ 0,\quad A=\Psi ,\ \Psi^* ,
\end{array}\right.
\ee
where ${\cal N}=4i(1-2(1+\tg )\vert\Psi\vert^2)/N$.  
Substituting these ${\cal C}_A^i$ and ${\cal B}_A^i$ into Eq.\ (\ref{heff})
reveals that
$H_{\rm eff}$ depends only on $\vert\Psi\vert^2$ and that its
extremum is at $\Psi =0=\Psi^*$. In the DIAG case
the actual value of the ground state energy density at this extremum is
given by:
\be\label{hamin}
\hat{H}\vert_{\rm min}=-\frac{2(1+\tg)}{\tl}(h_{\rm R}+h_{\rm L})^2\,,
\ee
and the expectation value of $f^\prime$ is:
$
f^\prime_0=-{4(1+\tg)}(h_{\rm R}+h_{\rm L})/{\tl}
$. We note that
$\hat{H}\vert_{\rm min}$ agrees (as it should) with the corresponding
result of the deformed $\sigma$-model 
with
$\tl\mapsto\lambda$, $\tg\mapsto g$ (Eq.\ (3.20) in \cite{BF}). 
For the RIGHT case the analogous expressions
of the axion model are simply obtained from
(\ref{hamin}) by setting $h_{\rm L}=0$.

At this point we recall the somewhat unusual
feature of the axion model once more, i.e.\ that the U(1) symmetry 
of the $S$-matrix (\ref{Smatrix}) is
realized through a topological current analogously to the Sine-Gordon
theory. To emphasize this we quote here the value of the classical free
energy density corresponding to
the Noether charge of the `manifest' U$_\theta$(1) symmetry of the Lagrangian
(\ref{Lagax}):
$\hat{H}\vert_{\rm min}^{(\theta)}=-{2h_{\rm R}^2}/{\tilde{\lambda}}$,
quite different from  Eq.\ (\ref{hamin}) with $h_{\rm L}=0$. 

To obtain the one loop correction to the free energy, one has to expand  
$\hat{\cal L}^d$ around the minimum, Eq.\ (\ref{hamin}). 
Writing $f=xf^\prime_0+\hat{f}$, where the expectation value of
$\hat{f}$ vanishes, one finds that the quadratic terms containing
$\hat{f}$ are independent of $h_{\rm L}$, $h_{\rm R}$, 
so the quadratic pieces of $\hat{\cal L}^d$
(hence the one loop correction)
are effectively determined by $\Psi=\sqrt{\tl/2}\,\psi$ only:
\be\label{Lagquadr}
{\cal L}_2=\pa_\mu\psi\pa^\mu\psi^*
-m^2\vert\psi\vert^2+iq(\psi\dot{\psi}^*-\psi^*\dot\psi )\,,
\ee
where $m^2=4h_{\rm L}(h_{\rm L}\tg_0+h_{\rm R}(1+\tg_0))$ and 
$q=h_{\rm L}(1-\tg_0)-h_{\rm R}(1+\tg_0)$. After continuation to
Euclidean space and writing
$\psi=(\phi_1+i\phi_2)/{\sqrt{2}}$, Eq.\ (\ref{Lagquadr})
becomes identical to the corresponding pieces for the 
deformed $\sigma$-model, Eq.\ (C.11) in \cite{BF} 
(with $\tl_0\mapsto\lambda_0$, $\tg_0\mapsto g_0$).
From this it follows that the free energy densities 
fully agree also for the DIAG and RIGHT cases in both models.

Now the results of the comparison 
between the 
free energies computed by the TBA
(based on the proposed $S$ matrix (\ref{Smatrix}))
and in PT for the deformed $\sigma$-model (\ref{Lagdef})
in Ref.\ \cite{BF} 
can be simply taken over for the axion model.
The conclusion is that there is complete consistency
between the TBA and the perturbative calculations,
providing good evidence for the validity of
the proposed $S$ matrix (\ref{Smatrix}) for the axion model.
Since the effective
coupling (\ref{betaeff1}) is identical in the two models
the $m/\Lambda_\MSb$ ratio found in \cite{BF} stays unchanged.

Finally we would like to point out that it would be 
interesting to study the axion model by lattice Monte-Carlo
simulations. This would provide us with a completely non-perturbative 
way of testing quantum T-duality. Technical difficulties arising
from the non-reality of the Euclidean action in the context of the
lattice Monte-Carlo study of the $O(3)$ model with a
constant $\theta$ term and a suggestion how to circumvent them
is discussed in \cite{Bietenholz}.

\noindent{\it Acknowledgements}
J. B. gratefully acknowledges a CNRS grant.
This investigation was supported in part by the Hungarian National
Science Fund (OTKA) under T~030099 and T~ 029802 and by the
Hungarian Ministry of Education under FKFP 0178/1999.


\begin{thebibliography}{99}
\bibitem{axions} For a review see J.E.~Kim, {\tt astro-ph/0002193} and
references therein. 
%
\bibitem{class}
J.~Balog, P.~Forg\'acs, Z.~Horv\'ath and L.~Palla,
Phys.~Lett.~{\bf 324B} (1994) 403.
%
\bibitem{bfhp}J.~Balog, P.~Forg\'acs, Z.~Horv\'ath, L.~Palla,  
  Nucl.~Phys.~B (Proc. Suppl.)
{\bf 49} (1996) 16. ({\tt hep-th/9601091})
%
\bibitem{Fateev}
V.A.~Fateev, Nucl. Phys. {\bf B473}[FS] (1996) 509.
\bibitem{BF}
J.~Balog and P.~Forg\'acs, Nucl. Phys. {\bf B570} (2000) 655.
({\tt hep-th/9906007})
%
\bibitem{Cherednik}
I.V.~Cherednik, Theor.~Math.~Phys.~{\bf 47} (1981) 422.
%
\bibitem{Kirillov}
A.~Kirillov, N.Yu.~Reshetikhin in Proc.\ of Proceedings of the Paris-Meudon
Colloquium,
String Theory, Quantum Cosmology and Quantum Gravity, Integrable
and Conformal Invariant Theories, (1986), eds.\ N.~Sanchez, H.~de Vega,
(World Scientific, Singapure).
%
\bibitem{SMpi}
I.~Affleck Field theory methods and critical phenomena, in
Fields, strings and critical phenomena, ed.\ E.~Br\'ezin and
J.~Zinn-Justin (North Holland, Amsterdam,1990);\\
V.A.~Fateev and Al.B.~Zamolodchikov,
Phys. Lett. {\bf 271B} (1991) 91;\\
A.B.~Zamolodchikov and Al.B.~Zamolodchikov, 
Nucl. Phys. {\bf B379} (1992) 602.
%
\bibitem{BN} 
J.~Balog and M.~Niedermaier,
Nucl. Phys. {\bf B500} (1997) 421.
%
\bibitem{Busch} T.H.~Buscher, Phys. Lett. {\bf B201} (1988) 466,
ibid. {\bf B194} (1987) 59.
%
\bibitem{Alvarez}E.~Alvarez, L.~Alvarez-Gaum\'e, Y.~Lozano,
Phys. Lett.  {\bf B336} (1994) 183.
%
\bibitem{alvtop} E.~Alvarez, L.~Alvarez-Gaum\'e, J.L.F.~Barb\'on and
Y.~Lozano, Nucl. Phys. {\bf B415} (1994) 71.
%
\bibitem{Byt}
A.G.~Bytsko,
{\tt hep-th/9403101}.
%
\bibitem{history}
A.~Polyakov, P.B.~Wiegmann, Phys. Lett. {\bf 131B} (1984) 121,\\
G.~Japaridze, A.~Nersesyan and P.~Wiegmann, Nucl. Phys. {\bf B230} (1984) 511,\\
P.~Hasenfratz, M.~Maggiore and F.~Niedermayer,
Phys. Lett. {\bf 245B} (1990) 522.
%
\bibitem{Osb}H.~Osborn, Ann. Phys. {\bf 200} (1990) 1.
%
\bibitem{Karp}
Z.~Horv\'ath, R.L.~Karp and L.~Palla,
{\tt hep-th/0001021}, to appear in PRD.
%
\bibitem{Bietenholz}
W. Bietenholz, A. Pochinsky and U. J. Wiese,
Nucl. Phys. Proc. Suppl. 47 (1996) 727.
%
\end{thebibliography}
\end{document}